# Development of a wearable haptic game interface

J. Foottit, D. Brown, S. Marks and A.M. Connor*

Auckland University of Technology, Colab (D-60), Private Bag 92006, Wellesley Street, Auckland 1142

**Abstract**

This paper outlines the development and evaluation of a wearable haptic game interface. The device differs from many traditional haptic feedback implementation in that it combines vibrotactile feedback with gesture based input, thus becoming a two way conduit between the user and the virtual environment. The device is intended to challenge what is considered an "interface" and sets out to purposefully blur the boundary between man and machine. This allows for a more immersive experience, and a user evaluation shows that the intuitive interface allows the user to *become* the aircraft that is controlled by the movements of the user's hand.

## 1. Introduction

Haptic technology [1] has found acceptance across a broad spectrum of fields, from highly sophisticated simulators for training surgeons to everyday devices like mobile phones. The diversity of applications also means there is a great diversity in the forms of implementation. Haptic feedback technology can be broadly divided into two main categories.

The first category consists of literal real-world simulations where the aim is to recreate the experience of touching real-world objects utilising an artificial interface [2]. This could include some adjustment of the real-world experience such as amplifying extremely small forces to make them perceptible to a human operator. The second category for haptic feedback technology is abstracted simulations where the haptic feedback provides information to the user that is not a literal representation of real-world forces. For example haptic feedback can be used to convey emotion [3], or to draw the attention of a user [4].

These two broad categories each have a range of implementations associated with them, however the first tends to rely on force reflecting interfaces that are capable of applying constant position-based forces on the user. Such implementations are relatively rigid, unwieldy and heavily steeped in mathematical rigor. In contrast, implementations in the second category vary considerably in terms of their size, scope and means for providing haptic feedback. When not attempting to simulate a real world event, force reflecting interfaces are rare. Force reflecting interfaces are capable of providing highly realistic representation of real-world experiences, however this fidelity comes at a cost, and often these interfaces are large, heavy and power consuming. These interfaces are utilized for high end applications where real world simulations are the objective, such as virtual simulators for training surgeons [5]. The high fidelity required by these applications justifies the high cost, weight and size of such interfaces. However there remains a wide range of opportunities for haptic technology to be applied where the limitations of force reflecting interfaces make their implementation infeasible. This is where the second category of haptic feedback technology emerges.

For many applications of haptic feedback true, high fidelity representations of contact forces are not possible due to the cost, weight, and size involved. This is particularly true where the haptic interface needs to be portable and wearable. Although the entire body as a whole communicates a great deal, the limb of choice for expression and exploration through gesture and touch is the hand. The anatomy of the hand makes it uniquely suited to such activities. The versatility and range of sensation available means the hand can interact with and experience the world better than any other part of the body. Hand gestures can even become a language in their own right, as in the case of sign language. Given the natural and fundamental nature of this mode of interaction, it makes sense to utilize the hands for interaction with machines. However while many machine interfaces utilize the hands for interaction, there is often a very clear divide between the hand and the machine interface. This places the interface as a physical intermediary between the human and the machine, requiring the human to translate their intentions into the interface language. In order to create a more intuitive interface, it becomes necessary to shift some of the translation work from the human to the machine.

---

*Corresponding author. Email:andrew.connor@aut.ac.nz

The goal of this research is to explore a way of interacting with machines that minimizes the layer of translation by focusing directly on the human hand as an input/output device. This is achieved through the use of a haptic feedback glove device that is designed to become an orthotic for the user. By minimizing the perception of an intermediary interface, the haptic feedback glove enables the user to more naturally and intuitively communicate with a computer by acting as a tangible interface. In particular, this development of the glove is focused in the first instance as an intuitive and useable game interface. The focus of this current paper is an evaluation of two instances of the haptic glove to determine the degree to which this has been achieved.

## 2. Background and related work

There has been a great deal of exploration regarding alternative interfaces for human machine interaction, particularly with the improvement of sensing devices and processing power of computer systems. Hand based interfaces are not entirely new, with examples dating back to the 1980s. For example, Zimmerman et al [6] describes the development of a hand gesture interface device that utilizes multiple sensors to track hand position and gestures. These early interfaces are often heavily reliant on inverse kinematic models embedded in software and often require extensive calibration before use. Whilst the interface may ultimately be quite effective, the need for calibration reduces the intuitive nature and emphasizes the fact that the device is an external object rather than an invisible intermediary.

In many cases the hands have been focused on as a primary means of interaction without the need for a specific device, for example the SmartSkin [7] system uses interactive surfaces that are sensitive to human hand and finger gestures. As with the early devices based approaches, there are limitations to these gesture approaches in that they lack the ability to provide any form of haptic feedback to the user and therefore do not necessarily provide any higher degree of engagement than traditional interfaces.

Attempts have been made to combine the advantages of device based approaches with the advantages of gesture based approaches. For example, the Charade system [8] utilizes a tethered glove that interprets finger positions and hand orientations in the context of a heavily scripted gesture language. Again, there is no implementation of tactile feedback to the user to indicate what events have been successfully interpreted. In contrast, other approaches have been developed that do provide such feedback but lack the ability to interact with a digital environment. For example, Frati & Prattichizzo describe a haptic feedback glove that responds to an avatar in a virtual environment where the hand position is tracked using a Kinect controller [9]. This is just one of many systems that utilise some form of camera for tracking, with many others discussed in the literature [10-12].

Very few attempts have been made to integrate gesture tracking into game controllers. Ionescu et al describe one such attempt [13] which involves gesture tracking for one hand whilst using a game controller in the other. Such approaches seem unwieldy and further emphasize that the controller is an external object, not an invisible intermediary between man and machine.

Outside of academic research, a number of commercial systems are in production. A novel approach to user interfaces focusing on the hand is an interface called Thumbles[†]. This takes the approach of using physical objects to represent virtual controls. The novel part of this approach is that physical artefacts change to suit the virtual environment. They move around and can be interacted with by the user to alter all kinds of controls. This emphasizes the importance of the human hand, but also focuses on the physical nature of an interface in preference to immaterial virtual interfaces. A haptic feedback glove has the potential to add a level of physicality to a virtual interface without needing to resort to physical objects that can become distracting and limited. Rather than altering the physical interface, the aim of haptic gloves is to make the physical interface invisible to the user, immersing them into an intuitive virtual environment that can be as dynamic and varied as the imagination allows.

The Myo[‡] is another example of a modern approach to human machine interaction. This device also focuses on the hands as a primary means of interaction. In this case, the device measures electrical signals to the muscles of the hand to allow for gesture based input. It is also capable of providing haptic feedback, however the location of the device on the arm removes the haptic feedback from where it is most relevant – the fingertips. While the low profile and light weight of the device makes it ideal for freedom of motion, a glove can offer these same advantages as well as providing a platform for haptic feedback at the point where people are accustomed to receiving it. A glove is also something that people are very familiar with, and so long as it does not impede motion of the hand it can quickly be forgotten that it is even being worn.

---

[†] http://www.pattenstudio.com/projects/thumbles/
[‡] https://www.thalmic.com/en/myo/

The range of interfaces emerging provides a range of options for users, each with their own benefits and limitations. In most cases, existing hand based interfaces either provide some form of haptic feedback or they provide some means of gesture tracking or user input. To our knowledge, it appears that there has been no attempt to provide both haptic feedback and gesture tracking in the same device, particularly when that device is intended to be an "invisible interface".

Different situations are likely to require different solutions, however the versatility of these emerging technologies means that more intuitive interaction with machines is becoming accessible in an ever increasing range of environments. The goal of this project is to develop a device that operates in this manner and provides a successful integration of different technologies.

## 3. Usability evaluation metrics
This paper outlines the ongoing development of a wearable haptic game interface. It expands previous work [14] by focusing on a more formal evaluation of the ability of the interface in use. To undertake the evaluation, a number of criteria were considered in advance of the evaluation taking place. In total, four different evaluation metrics were selected to determine the relative acceptability of the gloves. These are outlined in the following sections.

### 3.1. Accuracy and responsiveness
Accurate and responsive measurements of the hand and fingers are important since as soon as the user detects that their physical hand no longer corresponds to the digital representation the perception of direct control is lost. A users proprioception means that even if they cannot see their hand, they are still aware of the correlation between their physical hand motions and the virtual representation. This loss of direct connection requires the user to compensate for the inaccuracies of the system. The detriment of such a loss is twofold. Firstly, the user loses immersion as their focus is divided between the virtual experience and translating their desired input into a suitable motion for the input device. Secondly, the input device becomes unintuitive as the user must determine what physical motions correspond to what virtual motions. In contrast, if the hand and finger measurements are responsive and accurate the user need only think about the virtual representation of their hand and move their physical hand accordingly.

### 3.2. Freedom of motion
Freedom of motion is an important metric to consider for an immersive user experience. Early experiences during the development of the glove where the motion was restricted by the cable connection to the computer highlighted the importance of this metric. Without freedom of motion, the user is constantly required to be aware of the physical interface in order to keep track of how close to the limits they are. This distracts the user from the virtual experience by creating a dichotomous awareness split between the virtual experience and the physical interface. What is unclear is how free the motion must be in order to achieve the desired effect. For example, a long cable connection can provide an almost completely free range of motion that may be sufficient for most circumstances.

### 3.3. Comfort
In order to maintain the immersive experience for as long as possible, the glove must be comfortable. An uncomfortable device will distract the user from the virtual experience, particularly if the discomfort is acute enough for the user to describe it as painful. In contrast, if a device is comfortable the user can easily become accustomed to the presence of the device, even to the point that they are not specifically aware of it.

### 3.4. Robustness
The robustness metric is primarily a psychological measure of the perceived durability of the device. This is an important metric as it influences the way the user utilises the device. If the user is concerned about breaking the glove, then they are less likely to feel comfortable using it naturally. The impact on user experience is similar to a reduced freedom of motion, as it results in the user considering the limits of the device in a similar way to a restricted freedom of motion. The key difference is that in this case the freedom of motion is not limited by a physical constraint but a psychological constraint.

If the physical robustness of the device is not sufficient, then forces applied to the device by the user could cause it to malfunction or even stop working altogether. Any form of physical breakage causes a loss of perceived robustness, so an important aspect of this metric is ensuring the device does not physically break. However even without physical breakage, the design can influence the perceived robustness of the device. If the device does not appear robust it will still influence how it is used, potentially even to the same extent as if the device broke.

## 4. A wearable haptic glove

The overall goal of the project was to explore how to integrate technology into the human experience, with a particular focus on wearable haptic devices. It is a curious thing when a device becomes so natural that it is almost like an extension of the person. Whilst there are many advances in user interfaces that aim towards making the technology more natural and intuitive, achieving a truly integrated experience where the interface becomes part of a person's experience of themselves remains rare. At the forefront of this integration of technology and humanity is prosthetics and orthotics, technological devices designed specifically to integrate with the human body. Devices like mobile phones and cars that radically transform the way a person can interact with the world around them also tend to become integrated into the human experience over time, although these devices are much less likely to become a part of the user's perception of themselves the way an orthotic or prosthetic might.

A common thread emerges when looking at the technologies that successfully integrate into the human experience – the technologies must fit well with the human body. In our project, it quickly became clear that it would be vital for our glove to fit the hand comfortably and be light enough not to impede the mobility of the hand. Wireless communication was also important for our project, as being tethered to a computer creates a physical and psychological barrier that separates the technological device from the user's perception of themselves. A great deal of this change in the experience came from the need to keep track of the cable when using the wired solution. It distracted from the user experience by requiring them to be aware of the position of the cable in order to avoid it getting tangled or pulling out.

The key to wearable technology is effective miniaturization without losing features associated with larger devices. As devices become smaller and more energy efficient, it becomes possible to embed them into worn artefacts. In the case of providing input to a computer, a particularly useful piece of technology that has developed greatly in recent years is the Micro-Electro-Mechanical Systems (MEMS) based Inertial Measurement Unit (IMU). These remarkable units allow for the combination of sensors such as accelerometers, gyroscopes and magnetometers into incredibly small form factors and were a key aspect of the final glove design. In fact, these devices can be as small as a few millimetres in length, width and height. They are also increasingly affordable, and have become ubiquitous in mobile technology such as smart phones.

Although there has been much improvement in this area, these small, affordable devices are still considered somewhat inaccurate compared to their more expensive and bulky counterparts. However their accuracy is sufficient for most consumer applications, making them a glove-based input device. Their key limitation is a tendency to drift over time – particularly when being used to provide positional information. To overcome this limitation, technologies like the Microsoft Kinect can be used to supplement the data provided by these devices [9].

The development of the glove utilised a rapid prototyping methodology with various features trialled and refined to produce each of the final designs [14]. In particular, the prototyping involved considerable experimentation with different types of glove fabric, sensor and mounting for the haptic and other electronic components [15].

The flexibility, lightness and small size of the glove become an immediate focus, and using a custom fabricated PCB and a small form factor Arduino contributed a great deal towards achieving our goals in this area. The use of a custom knitted glove was also significant, as it was far more comfortable than the early prototypes. This was primarily due to the flexible nature of the fabric that still held the optical sensors in place. Again this was an iterated process and the comfort of the glove was a prime consideration during development.

This project required a wide range of support from different disciplines. The disciplines ranged from engineering to fashion to health sciences. This required interaction and coordination with people with very different sets of knowledge, each of which had something to contribute to the project. The role of the project team was to integrate the different aspects into a unified design, and to help each of the supporting people to understand enough of the other parts of the project to provide useful input. This bridging of disciplines to achieve a goal that could not be achieved independently is typical of students who excel in the Creative Technologies degree. Each student had their own leanings and preferences towards certain aspects of the project, but the greatest skills developed throughout this project were the ability to draw on the expertise of others to support the goals for the project.

## 4.1. Initial design

The first prototype used a MPU6050 MEMS inertial measurement unit to track hand orientation. This chip is manufactured by Invensense, and contains a 3-axis accelerometer and 3-axis gyroscope in a single package. Due to the small package size of the chip, a breakout board was used. The breakout board was connected to the main board of the glove by a standard 2.54 mm spaced header. The glove and corresponding electrical circuitry is shown in Figure 1.

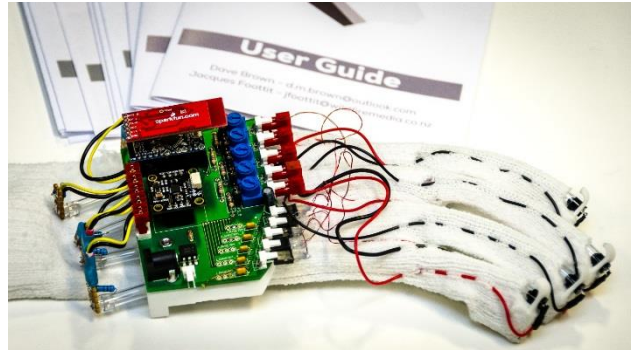

**Figure 1.** Initial glove design

Finger flex was measured using custom made optical bend sensors. These sensors were constructed using PVC tubing with an infra-red led at one end and an infra-red receiver at the other. The infra-red receiver was a simple light-dependent resistor sensitive to the infra-red frequencies of the infra-red LED. Thus as the tube was bent light attenuation resulted in an increase in resistance across the receiver, providing a flex measurement. The change in resistance was converted to an analog voltage input using a voltage divider circuit connected to the microcontroller.

Haptic feedback was provided using Eccentric Rotating Mass motors mounted in custom-made parts that were positioned at the finger tips. The custom made mounting parts doubled as end points for the optical bend sensors. The ERM motors were driven by Texas Instruments DRV2603 haptic motor driver chips. The motor drivers were mounted directly to the Haptic Glove main board, with standard 2.54mm headers providing connection points for the motors. The drivers were controlled using a Pulse Width Modulation signal from the microcontroller.

The microcontroller used in the first prototype was an Arduino Pro Mini from SparkFun. This board is based on the ATMEGA 328P, and provides an additional analog input to the standard Arduino Uno board. The microcontroller board provides a 5-pin interface for connecting it to the computer either using a FTDI interface with a USB cable, or a Bluetooth connection using the SparkFun Bluetooth modules. The board can only be programmed over USB using the FTDI connection, it is unable to be programmed over Bluetooth.

The first prototype operated at 5v, with the microcontroller and IMU board each having their own voltage regulators. The board was reliant on being provided with externally regulated power such as that provided by a USB connection.

## 4.2. Evaluation of initial design

User testing of the first prototype was conducted at a New Zealand Game Developers Association Meetup. A basic flight simulator game that was developed as part of the early phases of this work [14] was used to demonstrate the capabilities of the glove. The testing process was voluntary, and involved the user playing the flight simulator using the glove for as long as they desired. The user was then offered the option to provide their email address to receive an online survey. Video recording was also made of the users as they used the glove. Users who provided an email address were sent an email the following day with a link to the online survey. The online survey asked the users to rate the glove according to a number of metrics, as well as offering potential for written feedback. Users were also asked to rate the importance of a variety of features relevant to the glove's design in relation to a number of different evaluation metrics.

During the testing process, 7 users were filmed using the glove. 4 users provided email addresses, and 3 users responded to the email by completing the online survey. In addition, one user who was not filmed and did not provide an email address gave verbal feedback that was recorded.

The online survey utilised five point Likert scales and the mean responses are presented in Table 1. The mapping of each question to the four evaluation metrics discussed in section 3 are indicated in parenthesis.

Table 1. Summary of survey responses

| Question | Mean |
|---|---|
| The glove was easy to put on (C) | 3.00 |
| I was worried about damaging the glove while putting it on (R) | 3.67 |
| Putting on the glove was like putting on a regular glove (C) | 2.67 |
| The size of the glove made it difficult to put on (C) | 2.33 |
| I wasn't sure how to take the glove off (C) | 3.00 |
| The size of the glove made it difficult to take off (R) | 2.33 |
| The glove was easy to take off (C) | 2.00 |
| I was worried about damaging the glove while taking it off (R) | 4.67 |
| The glove was very responsive (A) | 3.00 |
| Controlling the aircraft was intuitive (F) | 4.33 |
| I noticed the aircraft started to turn to one side over time (A) | 4.00 |
| I found the delay on the finger triggers annoying (A) | 4.00 |
| I found it difficult to shoot targets (A) | 3.33 |
| It was easy to get the aircraft to go where I want (A) | 4.33 |
| It felt like my hand was the aircraft (F) | 4.33 |
| I noticed my hand getting tired quickly (C) | 4.33 |
| I got dizzy while playing the flight simulator game (C) | 2.33 |
| The glove was comfortable to wear (C) | 3.67 |

The survey also included options for the users to rank the importance of the desired features of the glove, again using a five point Likert scale. The responses are presented in Table 2.

Table 2. Desired features

| Question | Mean |
|---|---|
| Wireless | 3.67 |
| Haptic Feedback (vibration) | 3.67 |
| Accurate hand orientation | 4.33 |
| Accurate finger flex sensing | 4.67 |
| Hand position tracking | 4.00 |
| Low weight | 3.67 |
| Washable inner glove | 3.33 |
| Force feedback on the fingers | 2.67 |
| Small profile (not a bulky glove) | 2.00 |
| Small resistance to motion | 4.00 |

The data collected was used to inform the design decisions for a refined version of the glove. A primary focus of the design improvements was the use of Inertial Measurement Units or IMUs for finger flex sensing. In addition, a great deal of time was spent improving the integration of a digital compass into the wrist sensor to counteract IMU drift, an integrated battery and bluetooth module for wireless connectivity, and a built in battery charger to allow the integrated battery to be charged via a USB connection. These changes allowed the glove to be used without the use of a Kinect to track position and also freed the user from any form of tethering.

### 4.3 Refined design
The second prototype was designed to respond to user feedback provided during testing of the first prototype. Since a large portion of the cost involved in producing the prototype was in the PCB fabrication, as many features as possible were included in the PCB design although only a subset of these features were fully implemented in the current prototype. This was to allow for future expansion of the prototype without introducing additional costs related to additional PCB fabrication. The refined design for the glove is shown in Figure 2.

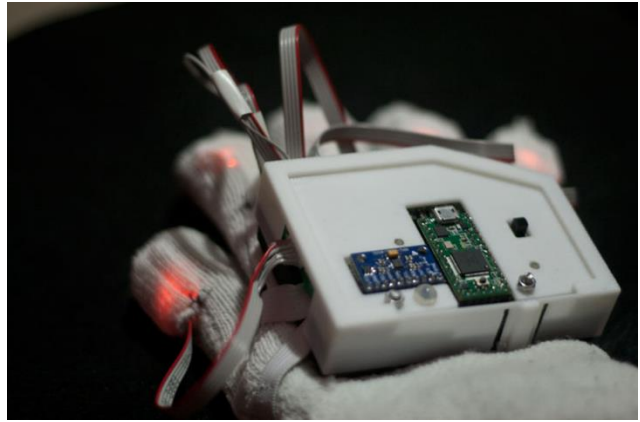

**Figure 2.** Refined glove design

The second prototype utilised Inertial Measurement Units for detecting the orientation of the user's fingers instead of using optical bend sensors. MPU6050 breakout boards were used for this purpose, providing 3-axis accelerometers and 3-axis gyroscopes in each sensor package. These sensors were mounted at the first distal Phalange and the second, third, fourth and fifth middle Phalanges. This arrangement provided complete orientation sensing for the thumb while removing the potential obstruction of finger movement for the remaining four fingers by keeping the sensors back from the tips of the fingers. The finger sensors were connected to the main glove board using ribbon cable that was directly soldered on to pads on the main board before being hot glued to provide strain relief. The thumb sensor was attached to the main board using a 5-pin JST plug with 2mm pitch spacing. In addition, the second prototype includes a connection for an additional arm-mounted sensor that could be used in the future to provide hand position information.

Hand orientation was measured using a MPU9250 sensor from Invensense. This sensor adds a magnetometer to the MPU6050 package, providing additional stability to the sensor by allowing further compensation for gyroscope drift. This sensor was also mounted on a breakout board, and was connected to the glove's main board by use of a standard 2.54mm spaced header.

Haptic feedback was provided using the same eccentric rotating mass motor configuration as the first prototype. The TI DRV2603 motor drivers used were also identical to the previous version of the glove. The connection between the motors and the glove was different in this version as a 10-pin JST plug with 1mm pitch spacing was used instead of standard 2.54mm headers.

The second prototype utilised a Teensy 3.1 microcontroller in favour of the Arduino Pro Mini. This microcontroller provided additional input capabilities as well as the ability to emulate HID devices such as a mouse. The Teensy microcontroller also allowed for the unit to be connected via Bluetooth and USB at the same time.

Bluetooth was included in the main board of the glove by the addition of a RN-42 module from Roving Networks. This surface mounted module allowed for Bluetooth communication between the glove and a computer and did not need to be disconnected in order to attach the microcontroller to the computer via USB.

A built-in Li-Po battery was included with second prototype, which also incorporated a 5v charger that was powered by the USB connection on the Teensy. This allowed the glove to be battery powered for more than 8 hours during testing, and easily recharged using a USB connection. Switching between operating mode and charge mode was achieved using a mechanical switch.

The second prototype operates a 3.3v, so two 3.3v buck-boost converters were included on the main board. These converters allow the 3.3v circuitry to operate throughout the discharge cycle of the on-board Li-Po battery. Due to the potential high current draw of the haptic feedback motors, the motor drivers were powered by a single buck-boost converter, while the remainder of the circuitry was powered by the second converter. A small tac switch was also included on the main board for the second prototype. This provides a programmable trigger for the glove, which could be used for functions such as resetting the default orientation of the glove.

## 4.4. Evaluation of refined design

User testing of the second prototype was conducted at the Auckland Armageddon Expo, a large entertainment expo. The user testing was voluntary, with users being permitted to try the glove for as long as they desired. Users were then given the option to supply an email address to be sent a link to online survey. There was also the option to fill out the online survey using a URL link if users preferred that to providing an email address. The testing process lasted for eight hours, with the glove being used for the entirety of that time except for a few minutes during the middle of the day where some quick repairs were made. Emails were sent out 2 days after the testing process, with a link to the online survey. The online survey was identical to the survey used for getting feedback on the first glove prototype.

While many people tested the glove, only six people provided email addresses. Some others took note of the URL option. To date only 1 person has filled out the online survey. Fortunately the person who has responded is a person who also tested the previous version of the glove, so was able to compare their experience with the first and second prototypes, however more data needs to be collected to support the ongoing evaluation of the glove. The online survey utilised five point Likert scales and the mean responses are presented in Table 3.

Table 3. Summary of survey responses

| Question | Mean |
|---|---|
| The glove was easy to put on (C) | 4.00 |
| I was worried about damaging the glove while putting it on (R) | 2.00 |
| Putting on the glove was like putting on a regular glove (C) | 4.00 |
| The size of the glove made it difficult to put on (C) | 3.00 |
| I wasn't sure how to take the glove off (C) | 3.00 |
| The size of the glove made it difficult to take off (R) | 2.00 |
| The glove was easy to take off (C) | 4.00 |
| I was worried about damaging the glove while taking it off (R) | 4.00 |
| The glove was very responsive (A) | 4.00 |
| Controlling the aircraft was intuitive (F) | 4.00 |
| I noticed the aircraft started to turn to one side over time (A) | 3.00 |
| I found the delay on the finger triggers annoying (A) | 2.00 |
| I found it difficult to shoot targets (A) | 3.00 |
| It was easy to get the aircraft to go where I want (A) | 4.00 |
| It felt like my hand was the aircraft (F) | 5.00 |
| I noticed my hand getting tired quickly (C) | 2.00 |
| I got dizzy while playing the flight simulator game (C) | 1.00 |
| The glove was comfortable to wear (C) | 5.00 |

## 4.5. Comparison

An important consideration in this work is whether the design decisions following the initial evaluation of the glove. This can be achieved by considering the responses of the survey questions in relation to the metrics outlined in section 3.

Figure 3 shows a visual representation of the survey data. A score for each of the four metrics has been determined by averaging the responses for each of the associated survey questions. A higher score indicates better performance for all of the metrics except robustness, for which a lower score is better.

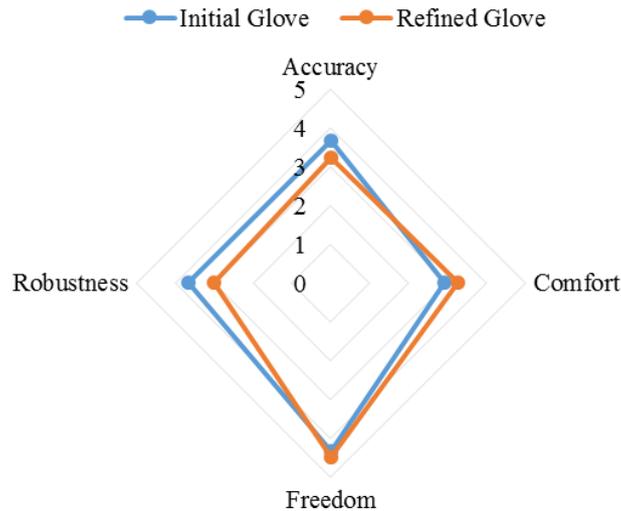

**Figure 3.** Evaluation comparison

The revised version of the glove can be seen to have improved performance when considered against 3 of the 4 metrics, but has resulted in a slightly reduced perception in terms of accuracy and responsiveness. This may in part be attributed to the decision to try and free the glove from the need for a Kinect to track the position of the glove.

## 5. Analysis and discussion
### 5.1. Considerations and limitations
The two user testing sessions had some limitations that need to be considered when analysing the results. One of the key limitations in comparing the results from the two surveys is the small sample sizes. The small number of responses to the online survey means it is difficult to make firm comparisons based solely on the data received. However, the insights obtained from the user testing remains very useful even with small sample sizes. In fact, Jakob Nielsen recommends no more than 5 users in a testing session [16]. Nielsen puts the emphasis on doing multiple, smaller tests rather than fewer larger ones, arguing that after about 4 users you have already discovered more than 75% of the usability issues. With this in mind, it is reasonable to draw some basic comparisons between the two prototypes despite the small sample size of the survey.

Another consideration that must be taken into account is the different locations of the user testing. While the first user testing session was held at a relatively small game developers meetup, the second user testing session was held at a large event that attracted large crowds. This resulted in a more diverse range of users at the second testing session, which resulted in problems being identified that may have been present but undetected in the first prototype. There was however one user that participated in both testing sessions, providing some valuable comparative data. In addition, the use of specific metrics limited the impact of identifying further usability problems.

### 5.2. Comparison of key metrics
During testing the first prototype exhibited very responsive hand measurements, however the optical flex sensors proved to be unreliable. This led to unresponsive and inaccurate readings of the finger flex. Despite these limitations, overall the users found the responsiveness of the glove to be adequate. It was clear that the lack of accuracy was the primary issue. Users reported both finding the delay on the fingers and the drift in orientation over time annoying. Both these phenomena were due to poor accuracy of the sensors utilised.

In the second prototype, the orientation drift was rectified by the addition of a magnetometer to the hand sensor. This provided almost completely drift free orientation sensing on the hand. In an attempt to improve the accuracy and responsiveness of the finger flex inertial measurement units were also used for the fingers, replacing the optical flex sensors. While users did not find issues with the accuracy of the hand orientation, the fingers remained troublesome. In addition, the responsiveness of the hand orientation was greatly reduced in the second prototype during testing. As a result, at the time of testing the second prototype did not have significant improvement in this metric. It was generally comparable to the first prototype, with improvements in hand accuracy and finger responsiveness being offset by a loss of responsiveness in the hand and loss of accuracy in the fingers. Despite these results, the limitations of the second prototype seem to be centred around the processing of data from the sensors, whereas the first prototype was limited by the sensors themselves.

One of the largest potential limitations identified for the first prototype was the freedom of motion. Due to the first prototype's reliance on external power and a USB connection, there was clearly a limit to the range of motion. In addition, the bulkiness of the optical flex sensors hampered wrist and finger movement slightly. During user testing however, these limitations did not pose significant issues for the users. In general the users appeared comfortable with their range of motion, and did not appear to limit their motion due to the presence of cables. In addition, the overwhelming user response indicated that the level of immersion was impressive, with a high level of agreement with the statement that they felt that their hand was the aircraft.

In order to improve the freedom of motion even further, the second prototype incorporated a built-in battery and Bluetooth connection with the computer. While it appeared there would be little room for improvement in this area, the second prototype did manage to demonstrate a superior freedom of motion. It was particularly noticeable when used by younger children – a demographic that was not present during the first testing session.

The comfort of the first prototype was not seen as a significant issue, with users reporting slightly favourable responses when asked about the comfort of the glove. However one thing that was clear was that the first glove caused the users hand to tire quickly, an observation that was backed up by the user reported data. In contrast, the second prototype appeared to fatigue users much more slowly. Even younger children were able to use the glove for periods exceeding 5 minutes, and often the use of the glove stopped to give another person a turn rather than due to fatigue. This was possibly due to the elimination of cables that added a downward force to the user hand, as well as the removal of the optical flex sensors that resisted finger flexion.

The robustness of the first prototype was a point of particular concern. Users frequently reported having concerns about damaging the glove when putting it on and taking it off. In addition, the USB and power cable connections to the glove were troublesome and did not provide a robust connection. This was somewhat mitigated for the user testing by securing the cables with cable ties, a solution that proved to be quite effective since users did not demonstrate specific concern about the cable connections and there were no instances of the connections coming apart during testing. Despite this improved robustness during testing users still felt concerned with the robustness of the prototype.

In an attempt to improve the perceived robustness for the second prototype, the main circuit board for the glove was mounted in a plastic case. In addition, each of the sensors were mounted in specially designed plastic cases. These cases improved the durability of the glove minimally, but added a perceived robustness not present in the previous prototype. This improved perception of robustness was clearly evident in the way users interacted with the device during the second user testing session. Users were much less cautious both when putting on and taking off the glove. This may have been influenced by the different demographic, with younger users particularly appearing more confident in the durability of the glove. This was an encouraging sign that the updated design provided a greater perception of robustness, however there was also two instances where cables were broken. This indicates that the actual robustness of the device did not match the perceived robustness. So while the physical robustness of the second prototype may have been similar to that of the first, the increase in perceived robustness means that the physical robustness needs to also be improved.

## 7. Future work
The overall goal of the project was to explore how to integrate technology into the human experience, with a particular focus on wearable haptic devices. The development work to date has utilised such a device as a game controller as a testing situation, and has aimed to produce an untethered wearable device that achieves satisfactory performance when measured against the evaluation criteria in section 3.

Having achieved such a device, future work will mainly be focused on deploying the glove in situations that are not game-related. In the first instance, the glove will be deployed in an immersive visualisation environment [17] intended to allow users to interact directly with scientific and engineering data. This environment is housed in a motion capture suite that allows a person to use a set of markers as a "mouse" to select and interact with a particular point in the three dimensional space. This selection is passive and the use of a wearable device with haptic feedback opens many opportunities for bi-directional interaction with the underlying data. The use of the glove in conjunction with an accurate motion capture suite will address any concerns over the loss of accuracy with the untethered glove. In addition, future work will also focus on further improvements to the glove and the conduct of more in-depth user-evaluations in a ranges of different usage scenarios.

## 8. Conclusions

Overall the second prototype demonstrated significant improvements in the robustness metric. There was also improvement to the comfort and freedom of motion metrics, although these improvements had a minor impact on the user experience. The sensor accuracy and responsiveness did not see noticeable improvements, with the increase in hand sensor accuracy being offset by a loss of responsiveness. The finger sensors between the two prototypes were similar in their limitations. Despite the lack of improvement in the sensor accuracy and responsiveness metric, the shift in limitation from sensor output to data processing means that the second prototype has the potential to achieve a greater score in this metric with a firmware upgrade.